# Investigating the effect of expected travel distance on individual descent speed in the stairwell with super long distance

Xingpeng Xu [1], Zhiming Fang [1,*], Rui Ye [1], Zhongyi Huang [1], Yao Lu [1]

[1]Business School, University of Shanghai for Science and Technology, Shanghai 200093, China

* Corresponding author. E-mail: zhmfang2015@163.com

**Abstract:** Currently, there is an increasing number of super high-rise buildings in urban cities, the issue of evacuation in emergencies from such buildings comes to the fore. An evacuation experiment was carried out by our group in Shanghai Tower [1], it was found that the evacuation speed of pedestrians evacuated from the 126$^{th}$ floor was always slower than that of those from the 117$^{th}$ floor. Therefore, we propose a hypothesis that the expected evacuation distance will affect pedestrians' movement speed. In order to verify our conjecture, we conduct an experiment in a 12-story office building, that is, to study whether there would be an influence and what kind of influence would be caused on speed by setting the evacuation distance for participants in advance. According to the results, we find that with the increase of expected evacuation distance, the movement speed of pedestrians will decrease, which confirms our hypothesis. At the same time, we give the relation between the increase rate of evacuation distance and the decrease rate of speed. It also can be found that with the increase of expected evacuation distance, the speed decrease rate of the male is greater than that for female. In addition, we study the effects of actual evacuation distance, gender, BMI on evacuation speed. Finally, we obtain the correlation between heart rate and speed during evacuation. The results in this paper are beneficial to the study of pedestrian evacuation in super high-rise buildings.

# 1. Introduction

In modern urban cities, there are a growing number of super high-rise buildings. The casualties and property losses caused by emergency incidents are more frequent in these buildings, so the safety of super high-rise buildings has also attracted people's attention because these buildings involve characteristics of large crowd size, multiple floors and multiple evacuation channels. In recent years, researchers have employed a variety of scientific methods to investigate the evacuation characteristics of super high-rise buildings in order to enhance the evacuation efficiency. Specifically, the effect of influential factors such as individual attribute (gender [2, 3], age [1, 4], BMI [2, 5], and physical condition [6], etc.), visibility [7-9], physical load [10], merging flow [11-14] and the geometry of stairwell [15, 16] have been investigated.

Chen et al. [2] studied the ascending speed of pedestrians in the stairwell of a 20-story building and the moving speed of males was obviously faster than that of females. It was shown that the ascent speeds of males keep at around 0.62 m/s, while the speeds of females keep at 0.52 m/s. To detect the effect of age on the escape ability in super high-rise buildings, an experiment conducted by our research group in Shanghai Tower [1] showed that the average velocity of younger people (age < 30) is $0.777 \pm 0.229$ m/s and the average velocity of the older people (age > 30) is 0.699 $0.669 \pm 0.186$ m/s. Whether a pedestrian is disabled or not is also a vital factor that must be considered in the evacuation process of super high-rise buildings. Due to fact that the disabled may fail to travel through stairs smoothly, the evacuation time of them was longer than that of those without disabilities and the disabled significantly increased the risk of congestion and blocking on stair platforms [6, 17]. Zhang et al. [7] examined the effect of visibility on individual and group evacuation speed, and found that the evacuation speed of group was faster than that of individuals on stairs in lower visibility. Besides, In the emergency evacuation of stairwells, the queuing behavior, lag phase and merging flow tended to happen at stair platforms [9]. It was also found that heavy loads can affect the evacuation speed in the stairwell [10]. During this experiment, the group evacuation speed of each floor is 0.97m/s ~ 1.42 m/s, 0.78 m/s ~ 0.95 m/s and 0.68m/s ~ 0.95 m/s in the load conditions of 0 kg, 5kg and 10 kg. Furthermore, the difference in geometry for stairwells also affects the evacuation speed. Frantzich et al. [15] conducted experiments on two types of stairs and found that the average ascending and descending speeds of narrow stairs were 0.51 m/s and 0.72 m/s respectively, and the average ascending and descending speeds of wide stairs were 0.56 m/s and 0.69 m/s respectively. For instance, Chen et al. [18] conducted a series of single-file experiments to discuss the relation between density and velocity both in the descending and ascending direction. An empirical formula for calculating the expected speed of descending stairs was proposed base on evacuation experiment of a 21-story building [19, 20]. What's more, as a matter of fact, no matter under normal circumstances or emergency conditions, pedestrians tend to follow their familiar friends and relatives on crowd flow to find evacuation routes and exits [13, 21] and evacuate in small groups [22].

With respect to the evacuation efficiency in super high-rise buildings, establishment and application of evacuation models are common methods of investigating the evacuation characteristics in super high-rise buildings. Common models of pedestrian evacuation are divided into continuous model and discrete model, the former mainly includes social force model [23], the latter mainly includes cellular automata model [24-26] and lattice gas model [27], etc. For instance, Li et al. [23] took into account the unevenness of the ground, and proposed an extended social force model to reproduce the temporal-spatial pedestrian dynamics on the stairway. Unlike continuous models, the cellular automata model with the characteristics of small computation can discretize the staircase scene and regard pedestrians as cells, which has been widely employed in the evolution of the flow of people up the two-dimensional plane [28].

However, there is still a lack of experimental data of evacuation in super high-rise buildings at present, and the influence of distance on pedestrian evacuation speed during the process of long-distance evacuation has not been known yet. In 2017, our research group carried out an evacuation experiment with a vertical evacuation distance of 580m in Shanghai Tower, which is the second tallest building in the world [1]. One of the significant results is that we found that pedestrians move more slowly when they evacuated from higher floors, as shown in Fig. 1. We assumed that with the change of the expected evacuation distance, pedestrians may set appropriate evacuation speeds for themselves, that is, the longer the expected evacuation distance, the lower the evacuation speed, which may be related to the psychological expectations of pedestrians. In order to verify the validity of our hypothesis, we conducted another single-person experiment. In this experiment, we set different evacuation distance for participants to measure their actual evacuation speed. The experimental results found that expected evacuation distance does have an impact on evacuation speed. In this paper, we will describe the results of this experiment in detail.

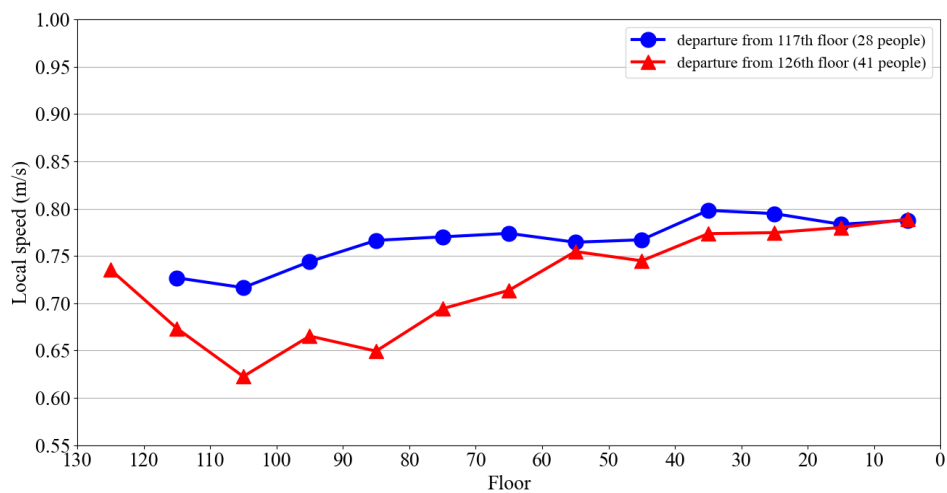

**Fig. 1.** Two groups evacuated from the 117$^{th}$ floor and 126$^{th}$ floor in Shanghai Tower.

It is worth noting that the influential factors and previous studies mentioned above mainly focused on physical layer and group behaviors. Even if some strategies has been discussed [29, 30] , researchers may ignore the impact of psychological expectations on long distance in the super high-rise buildings. Therefore, it is of practical significance to study the impact of pedestrian psychological expectation on evacuation. What's more, our research will contribute to the design of more scientific and reasonable pedestrian evacuation strategies in super high-rise buildings. The structure of this paper is as follows: Section 2 is the experiment carried out. The following Section 3 presents the measurement method and results are given in Section 4. The discussion is in Section 5. Finally, Section 6 summarizes our work.

## 2. Experiment

The experiment was carried out in a 12-story university office building in Shanghai, which has 248 steps, with an average of 22 steps between adjacent floors. It is a typical office building with a stairwell and an elevator that connect the 12 floors. A total of 42 participants who are all undergraduate and graduate students were recruited, including 21 males and 21 females between the ages of 22 to 28 years old. The experiment was performed on a clear day with a camera installed above the stair platforms each floor that could capture the pedestrian movements on stairs, as shown in Fig.2 (a). In order to record the heart rate of pedestrians when they evacuated the stairwell, each participant had to wear a sports bracelet during the whole process of experiment. The width (Fig. 2 (b)), riser and trend of most steps (Fig.2 (c)) are about 1.420m, 0.165m and 0.280m respectively. In other words, if the riser of all steps represent the height of this office building, the vertical height reaches 40.920m, which can meet the height requirement of high-rise buildings.

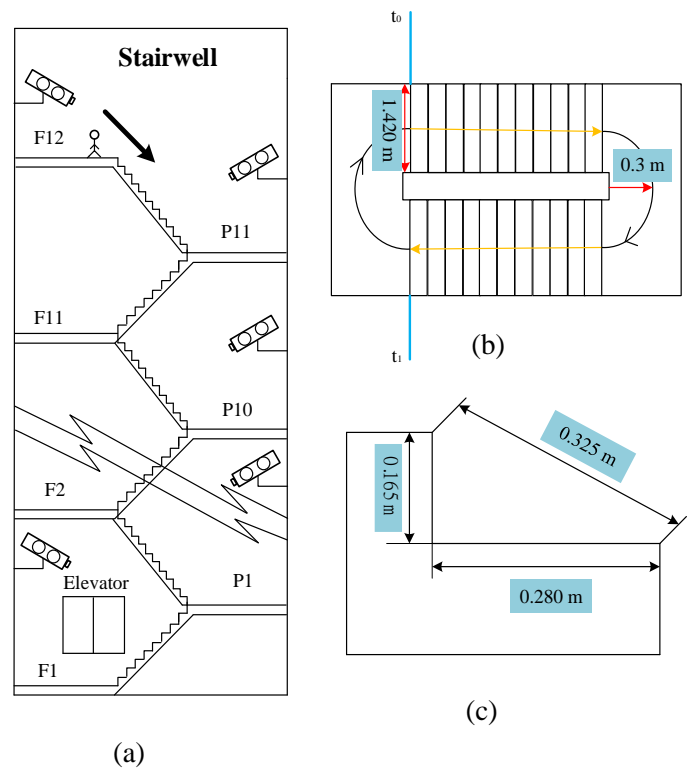

**Fig. 2.** Dimensions of the staircase: (a) section diagram of the stairwell, (b) schematic illustration of the stairs, (c) size diagram of the most steps.

To mimic the evacuation in super high-rise buildings, we ask each participant to move downward on stairs for multiple turns at a time in this 12-story building, and here one turn represents the process where certain pedestrian evacuates from the 12$^{th}$ floor to the ground floor, and then takes elevator to go back to the 12th floor. Furthermore, we define the process when certain pedestrian evacuates on stairs from the 12th floor to the ground floor as one lap. In order to better explore the impact of expected travel distance on descent speed, we design 7 experimental groups. For the sake of understanding the experiment setup, the 7 groups were named as G_5+3, G_6+2, G_7+1, G_8, G_9, G_10 and G_11, as shown in Table 1, where G_5+3 represents that the pedestrian was informed to run 5 turns before setting off, and he/she was told to run another 3 turns when he/she completed 5 turns later; G_8 means that the pedestrian was told to run 8 turns without other instructions during his/her evacuation process. 42 participants were randomly divided into 7 groups corresponding to the above 7 evacuation scenarios, specifically, 3 males and 3 females in each group. To make the experiment results more reliable, participants who came from different majors and schools were independent of each other, everyone did not know each other and had no idea how many turns he/she was going to run.

Table 1 Group description in different scenarios.

| Scenario | | Description | male | female |
|---|---|---|---|---|
| Number | Label | | | |
| 1 | G_5+3 | Initially told to run 5 turns, 5 turns later, 3 more turns | 3 | 3 |
| 2 | G_6+2 | Initially told to run 6 turns, 6 turns later, 2 more turns | 3 | 3 |
| 3 | G_7+1 | Initially told to run 7 turns, 7 turns later, 1 more turns | 3 | 3 |
| 4 | G_8 | Initially told to run 8 turns | 3 | 3 |
| 5 | G_9 | Initially told to run 9 turns | 3 | 3 |
| 6 | G_10 | Initially told to run 10 turns | 3 | 3 |
| 7 | G_11 | Initially told to run 11 turns | 3 | 3 |
| | | total | 21 | 21 |

# 3. Measurement method

## 3.1 Local speed

In our experiment, we mainly focus on the local speed, which is defined as the distance between adjacent floors, divided by the time lapse between the participant leaves and enter certain floor. In this building, the size of the steps that connect the platform on each flight of stairs varies slightly between floors through filed measurement. As shown in Table 2, the height of the most steps is represented by $S_{rm}$ and the number of most steps with the same height is represented by $n_{sm}$. The height of the unique step that reaches the platform is $S_{r1}$, and $S_t$ is the value of the tread surface of the stairs. F12_P11 represents that pedestrian downward from 12$^{th}$ floor (F) to the 11$^{th}$ platform (P) that connects two flights between 12$^{th}$ floor and 11$^{th}$ floor. Then, $ls$ is used to represent the distance of the stepped ramp as follows:

$$ls = n_{sm}\sqrt{s_t^2 + s_{rm}^2} + \sqrt{s_t^2 + s_{r1}^2} \tag{1}$$

The experimental phenomenon shows that the motion trajectory of pedestrians on the platform is usually a semicircle. For accuracy, a semicircle with a radius of 0.3m is set here. Then, the motion distance $l_t$ of pedestrians on the platform can be expressed by the formula:

$$l_t = 0.3\pi \approx 0.942 \text{m} \tag{2}$$

Therefore, the motion distance between floors ($\triangle d$) can be expressed by the following formula:

$$\triangle d = 2l_s + l_t \tag{3}$$

Table 2 Stairwell step size and movement distance considered in this study

|  | $n_{sm}$ | $S_{rm}$ (m) | $S_{r1}$ (m) | $S_t$ (m) | ls (m) | $l_t$ (m) |
|---|---|---|---|---|---|---|
| F12_P11 | 10 | 0.165 | 0.170 | 0.280 | 3.578 | 0.942 |
| P11_F11 | 10 | 0.165 | 0.155 | 0.280 | 3.570 | |
| F11_P10 | 10 | 0.165 | 0.165 | 0.280 | 3.575 | 0.942 |
| P10_F10 | 10 | 0.165 | 0.155 | 0.280 | 3.570 | |
| F10_P9 | 10 | 0.165 | 0.155 | 0.280 | 3.570 | 0.942 |
| P9_F9 | 10 | 0.165 | 0.165 | 0.280 | 3.575 | |
| F9_P8 | 10 | 0.165 | 0.165 | 0.280 | 3.575 | 0.942 |
| P8_F8 | 10 | 0.165 | 0.165 | 0.280 | 3.575 | |
| F8_P7 | 10 | 0.165 | 0.155 | 0.280 | 3.570 | 0.942 |
| P7_F7 | 10 | 0.165 | 0.165 | 0.280 | 3.575 | |
| F7_P6 | 10 | 0.165 | 0.165 | 0.280 | 3.575 | 0.942 |
| P6_F6 | 10 | 0.165 | 0.165 | 0.280 | 3.575 | |
| F6_P5 | 10 | 0.165 | 0.168 | 0.280 | 3.577 | 0.942 |
| P5_F5 | 10 | 0.165 | 0.165 | 0.280 | 3.575 | |
| F5_P4 | 10 | 0.165 | 0.155 | 0.280 | 3.570 | 0.942 |
| P4_F4 | 10 | 0.165 | 0.160 | 0.280 | 3.572 | |
| F4_P3 | 10 | 0.165 | 0.155 | 0.280 | 3.570 | 0.942 |
| P3_F3 | 10 | 0.165 | 0.16 | 0.280 | 3.572 | |
| F3_P2 | 12 | 0.165 | 0.162 | 0.280 | 4.223 | 0.942 |
| P2_F2 | 11 | 0.165 | 0.161 | 0.280 | 3.898 | |
| F2_P1 | 12 | 0.165 | 0.175 | 0.280 | 4.230 | 0.942 |
| P1_F1 | 11 | 0.165 | 0.154 | 0.280 | 3.895 | |
| total | | | | | 80.566 | 10.362 |
| | | | | | 90.928 | |

As mentioned earlier, above each platform there is a camera that can record the leaving and reaching time of certain stair section by certain pedestrian. Professional video player software is used to read the surveillance video of pedestrian movement on stairs. The camera on the previous platform can record the reaching time ($t_0$), and camera on this platform can record the leaving time ($t_1$). For example, the snapshot in Fig. 3(a) shows the time when the female enters the floor section F11-F10, and the one in Fig. 3(b) shows the time when she leaves F11-F10. Besides, since the frame rate of the video is 25 frames per second, the pedestrian's movement time in the stairwell can be accurately measured in milliseconds, which also ensures the accuracy of the results in the following data analysis. Then the evacuation time for pedestrians between adjacent floors is formulated by $\triangle t = t_1 - t_0$. The average speed of each participant between adjacent floors was defined as local speed, which was calculated as follows:

$$v_t = \frac{\triangle d}{\triangle t} \tag{4}$$

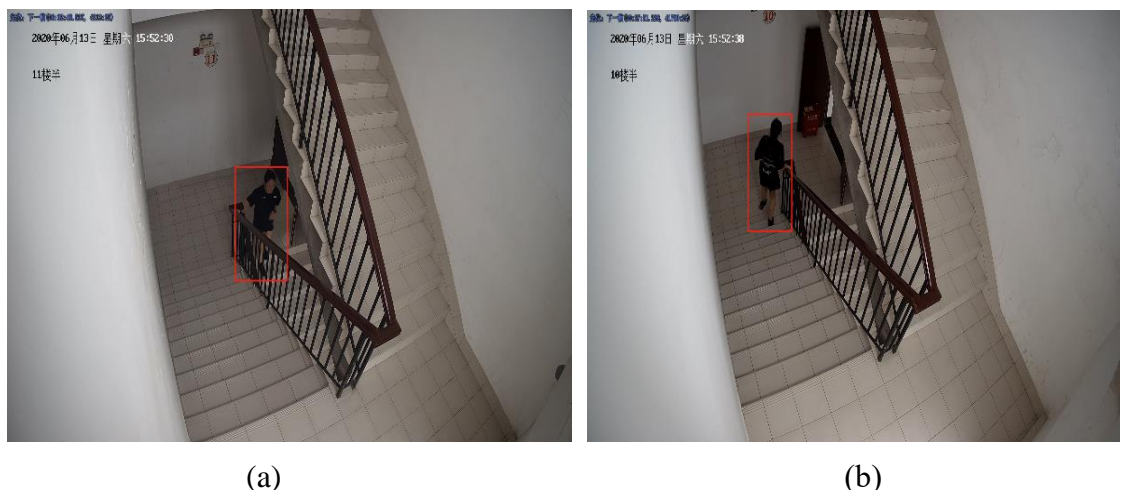

(a)                  (b)

**Fig. 3.** Snapshot of pedestrian evacuation: (a) the female was just leaving the 11[th] floor; (b) the female was just reaching the platform on 10[th] floor.

## 3.2 Heart rate

During the experiment, each participant wore a Garmin exercise bracelet, which could record not only the movement time but also the heart rate corresponding to the time. Through video monitoring, the reaching time ($t_0$) and leaving time ($t_1$) can be extracted, so the average heart rate within $\triangle t$ can also be obtained by unifying the time recorded by bracelets and the camera video time.

# 4. Results and analysis

## 4.1 Distribution of local speed

As mentioned above, we can obtain a local speed data point for adjacent floors. In other words, for a certain participant, 11 local speed points can be obtained in each of his/her laps. Specifically, there are 3 males and 3 females in G_5+3, 528 ($11 \times 8 \times 6$) local speed points can be obtained after 8 laps. The calculation method of the amount of data points in other groups is similar, so 4092 local speed points can be extracted from 7 different evacuation groups.

First of all, we show the distribution of 4092 local speed points under the premise of ignoring the difference of expected evacuation distance of each group. In order to compare gender differences, the speed distribution results of males and females are also given. For the above three distributions, we respectively use normal functions to fit them, and the results are shown in Fig. 4. The average movement speed of males and females is $1.26 \pm 0.37$m/s and $0.86 \pm 0.19$m/s, respectively, and the average speed of the whole is between that of males and the females. The standard deviation of male speed is greater than that of female. However, it's almost consistent for male speed and overall speed.

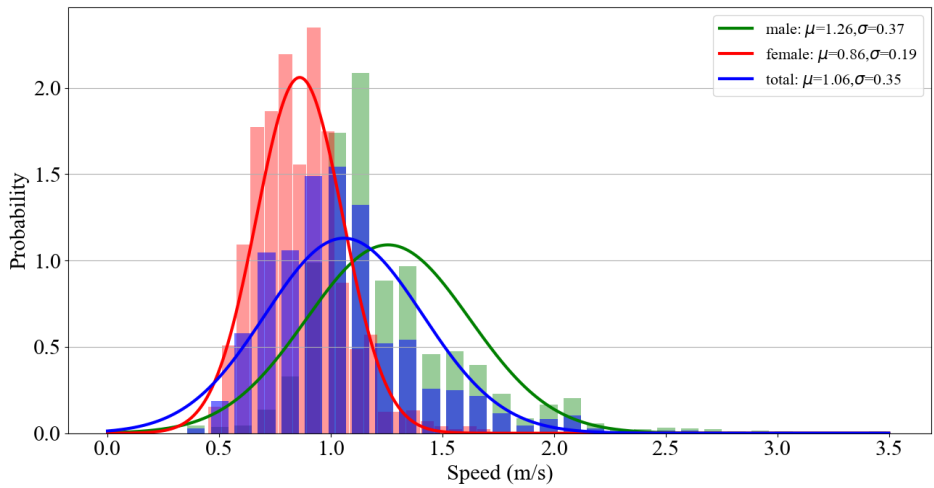

**Fig. 4.** The speed distribution of the whole, males and females.

## 4.2 Speed of adjacent floors and adjacent laps

The 4092 local speed points are classified according to the movement scene in adjacent floors, for example, all local speed points between $12^{th}$ floor and $11^{th}$ floor are classified as F12_F11. In order to study the local speed variation trend in adjacent floors, violin diagram and box diagram are used to represent the probability density and distribution characteristics of the local speed points in adjacent floors respectively (Fig.5–Fig.7). In the box diagram, the orange horizontal line represents the median value and the blue points represents the mean value. Furthermore, the average local speed of the whole, namely 21 males and 21 females is 1.18m/s, 1.40m/s and 0.97m/s in F12_F11, but they are 0.94m/s, 1.17 m/s and 0.78 m/s in F2_F1, respectively. We define F12_F11 as the $1^{st}$ movement scene and F2_F1 as the $11^{th}$ movement scene. Then the independent variable $x$ is the ordinal number of the evacuation scene of adjacent floors, and the dependent variable $y$ is the average evacuation speed. The downward trend from the $12^{th}$ floor to the ground floor is expressed as a red linear equation (Table 3). If we focus on the slope of linear functions, it can be found the speed of males between adjacent floors drops faster than that of females.

**Table 3** Fitting function between average speeds and floors

| | | 11 laps fitting plot | $R^2$ |
|---|---|---|---|
| evacuation speed between adjacent floors | all participants | $y = -0.027x + 1.245$ | 0.549 |
| | 21 males | $y = -0.021x + 1.383$ | 0.602 |
| | 21 females | $y = -0.015x + 0.956$ | 0.691 |

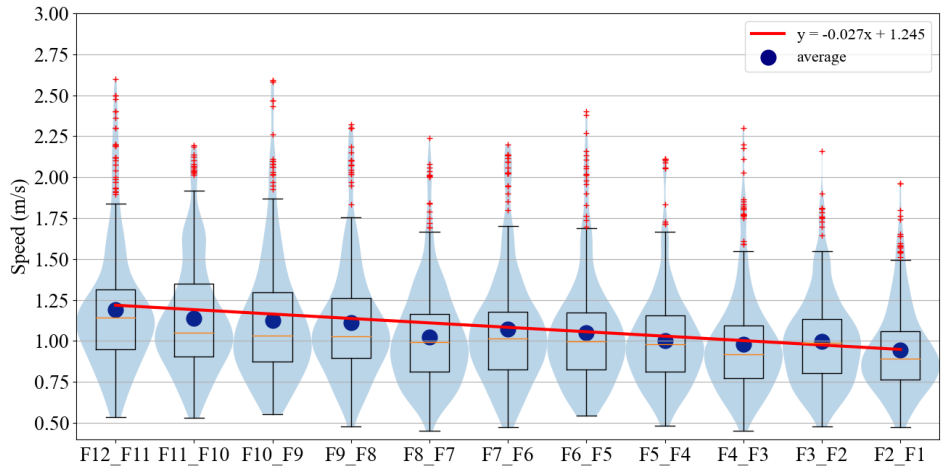

**Fig. 5.** The speeds of the whole between adjacent floors.

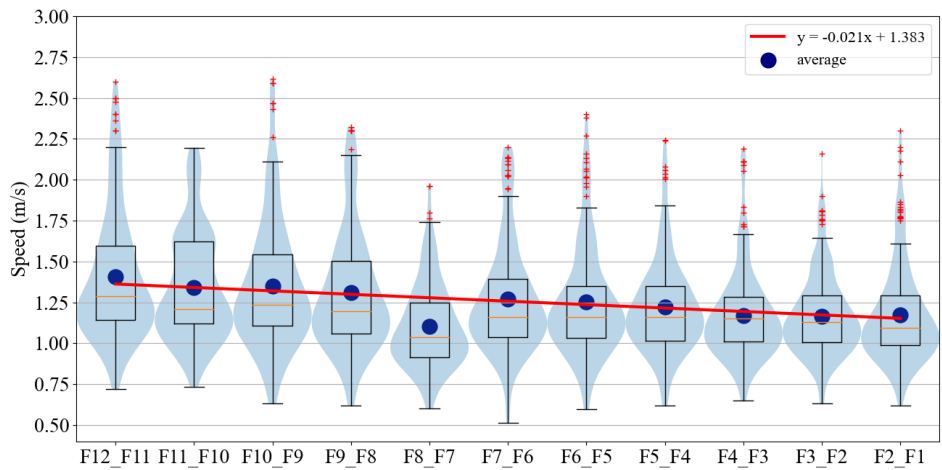

**Fig. 6.** The speeds of 21 males between adjacent floors.

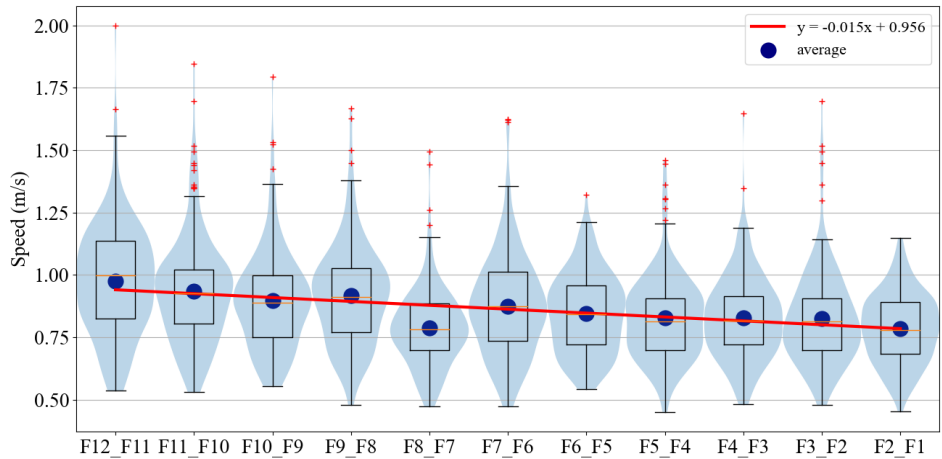

**Fig. 7.** The speeds of 21 females between adjacent floors.

In this evacuation experiment, pedestrians descend at least 96 floors (8 laps) and take an elevator to the 12th floor after reaching the ground floor. In this 12-story building, it takes $33.50 \pm 0.18s$ to get from the ground floor to the $12^{th}$ floor by elevator. The time to take the elevator can be regarded as the time to have a rest. It is not difficult to find that the speed of the pedestrians increased significantly after each lap with a break (Table 4). After the rest, the speed of male, female and all participants increased by 28.0%, 28.5% and 27.7%, respectively. In order to ensure safety, the refuge floor is often set up to provide temporary refuge or rest for the crowd. So the role of taking the elevator here is actually similar to taking a break on refuge floors.

**Table 4** Effects of rest on evacuation speed.

|  | f2_f1 | | | f12_f11 | | |
|---|---|---|---|---|---|---|
|  | all participants | 21 males | 21 females | all participants | 21 males | 21 females |
| median (m/s) | 0.89 | 1.09 | 0.77 | 1.14 | 1.28 | 0.99 |
| average speed (m/s) | 0.94±0.26 | 1.17±0.30 | 0.78±0.13 | 1.18±0.38 | 1.40±0.39 | 0.97±0.22 |

Then, the 4092 local speed points were classified according to the lap number. From the $1^{st}$ lap to the $11^{th}$ lap, the average speed of each lap has a downward trend (Fig. 8). Since the pedestrian in the first lap is in the best physical fitness and has the lowest sense of fatigue, compared with other laps, there are some speed data points whose values are larger than 2.25 m/s.

In order to study the influence of gender, the local speed points of 21 males and 21 females were extracted, as shown in Fig. 9-Fig. 10, the male participants evacuated much faster than the female participants on the stairs. Also, with the increase of movement distance, the evacuation speed will be smaller for both male and female participants. Besides, here the independent variable $x$ is the lap number, and the dependent variable $y$ is average speed. It can be found that the linear fitting of the mean values of the speed shows that the downward trend of first 8 laps (black line) is slightly higher than that of 11 laps (red line). Similar to the results in Table 3, if we focus on the slope of linear functions in Table 5, the speed of male decreased more rapidly than that of female with the increase of evacuation distance.

**Table 5** Fitting function between average speeds and laps.

|  |  | 8 laps fitting plot | $R^2$ | 11 laps fitting plot | $R^2$ |
|---|---|---|---|---|---|
| evacuation speed model | all participants | $y = -0.032x + 1.193$ | 0.923 | $y = -0.025x + 1.167$ | 0.901 |
|  | 21 males | $y = -0.039x + 1.447$ | 0.921 | $y = -0.036x + 1.436$ | 0.950 |
|  | 21 females | $y = -0.024x + 0.969$ | 0.906 | $y = -0.021x + 0.951$ | 0.914 |

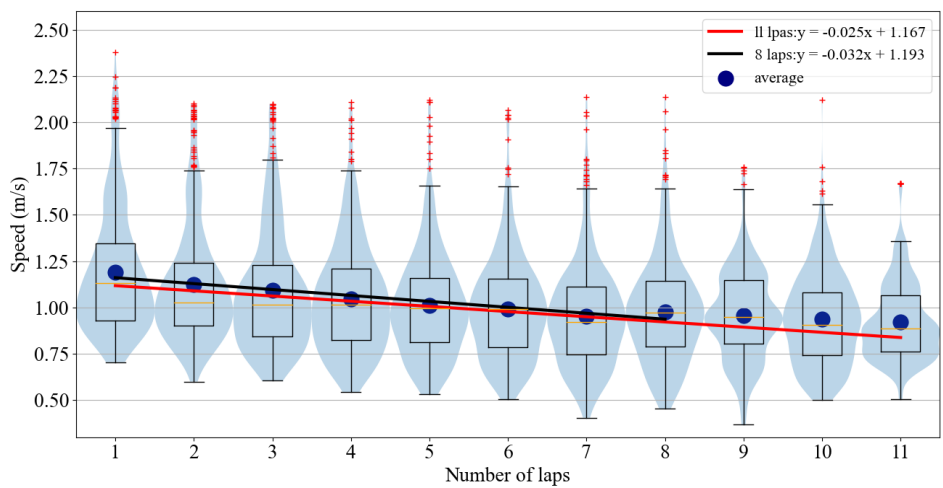

**Fig. 8.** The speeds of the whole per lap.

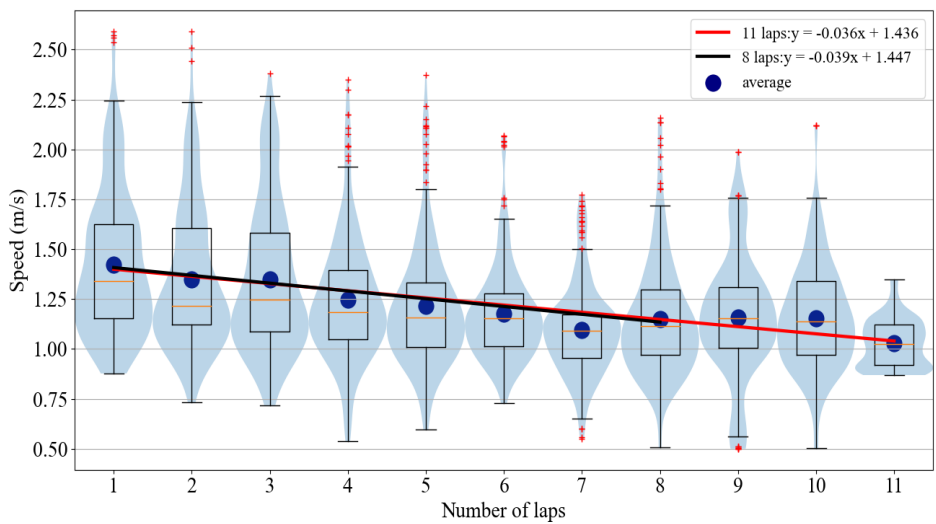

**Fig. 9.** The speeds of 21 males per lap.

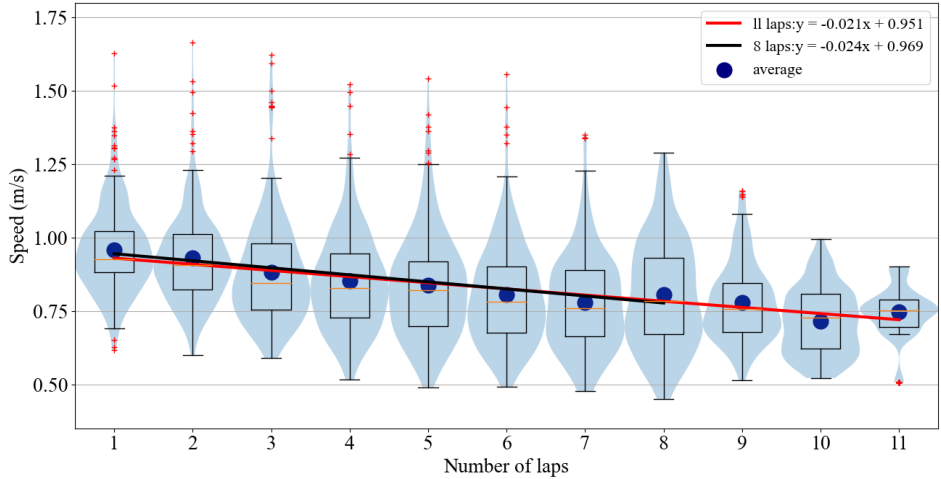

**Fig. 10.** The speeds of 21 females per lap.

## 4.3 The effects of psychological expectation on evacuation speed

In this section, we study the effect of pedestrians' expected evacuation distance (i.e., expected evacuation laps) on evacuation speed on stairs in super high-rise building. To avoid the effects of outliers on psychological expectations, Three-Sigma rules are used to filter the abnormal speed points in different situations before analysis. On the basis of Fig. 8-Fig. 10, the relations between average speed of 7 groups (Table 1) and expected evacuation distance are presented in Fig. 11-Fig. 13. The results are as follows:

At the beginning, the initial instructions of 7 groups required participants to complete 5 to 11 laps respectively (Table 1). The initial instruction makes a significant difference in movement speed during the first lap of each group (Fig.11-Fig.13). Due to the fact that participants were required to evacuate different number of laps before departure, not only the speed of 7 groups differ significantly in the 1$^{st}$ lap, but also the speed of subsequent lap decreases as the number of laps increases. So, on the whole, G_5+3 has the highest speed and G_11 has the lowest speed in Fig. 11-Fig. 13. Furthermore, the maximum speed of G_5+3 is 1.30m/s in the first lap, while the evacuation speed of G_11 is 1.09 m/s. In other words, the expected evacuation distance will have an effect on pedestrians' movement speed and the longer the evacuation distance is, the lower the "appropriate speed" set by pedestrians themselves will be. In addition, it can be found that in the group with only 8 evacuation laps, the speed of participants would suddenly increase in the next lap after receiving the additional evacuation number instruction, such as the 6$^{th}$ lap in G_5+3.

Besides, in 7 groups of evacuation scenarios, the evacuation speed of females is obviously slower than that of males (Fig. 12-Fig. 13). From the perspective of wholeness, the male speed also fluctuates greater than female, which makes the evacuate speed of the female more regular than the male.

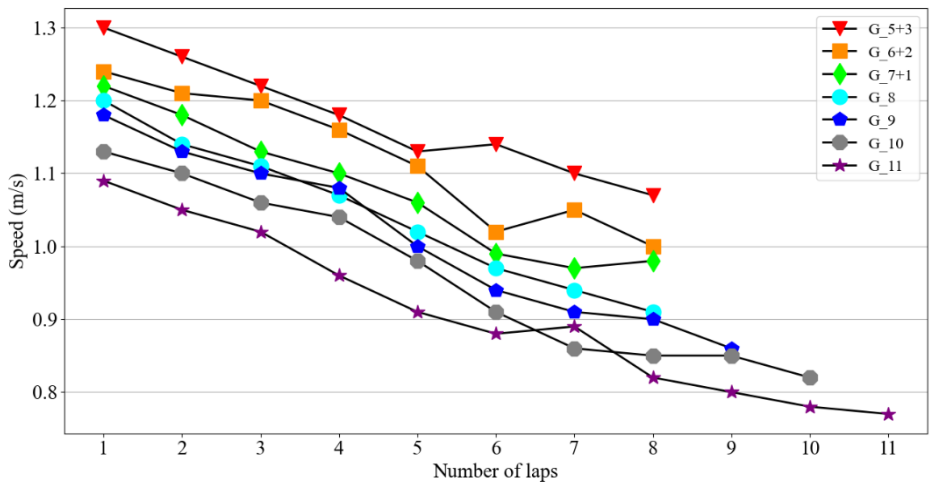

**Fig. 11.** The relation between average speed and evacuation laps in all 7 groups

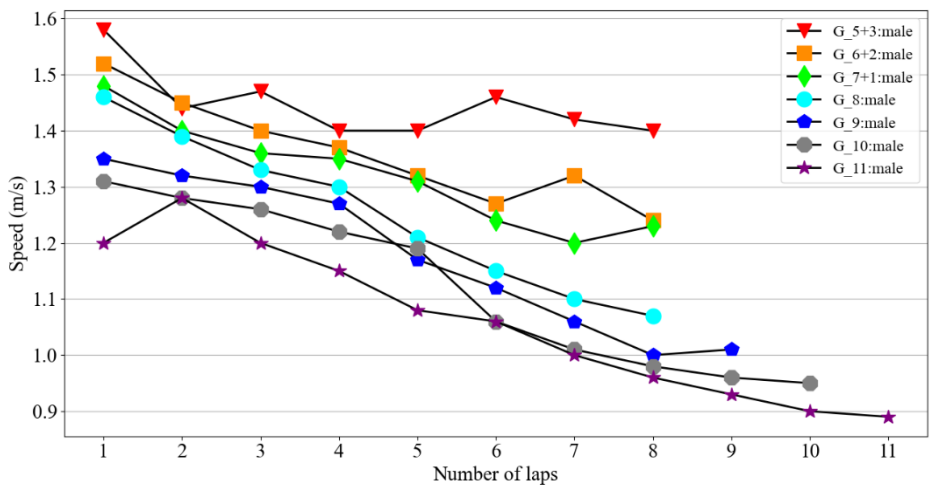

**Fig. 12.** The relation between male average speed and evacuation laps in all 7 groups.

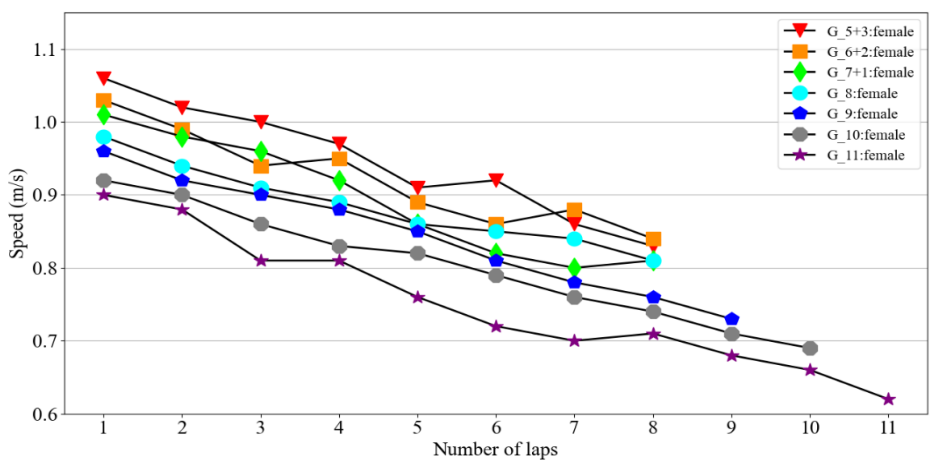

**Fig. 13.** The relation between female average speed and evacuation laps in all 7 groups.

In addition, the female speed of G_5+3, G_6+2, G_7+1 and G_8 are 32.65%, 31.91%, 31.15% and 31.34% slower than male speed respectively in the first 5 laps. However, the average female speed of G_9, G_10 and G_11 in first 5 laps are 29.68%, 29.83% and 29.66% slower than the males (Table 6). In other words, when the expected evacuation distance is greater, the difference in evacuation speed between male and female will be smaller. To further explore the influence of psychological expectation on evacuation speed, for different experimental groups, with G_5+3 as the control group, we obtain the relation between increase rate of expected evacuation distance and decrease rate of evacuation speed in Fig. 14. Taking G_5+3 and G_6+2 as an example, the expected evacuation distance is 5 and 6 laps respectively, and the average speed of the first 5 laps in these two groups is 1.204m/s and 1.174m/s respectively. Therefore, when the expected evacuation distance increases by 20%, the evacuation speed decreases by 2.49%. Besides, according to the fitting results, the expected evacuation distance has a greater impact on the male.

Table 6 Statistics of speeds in first 5 laps in all 7 groups.

| | | G_5+3 | G_6+2 | G_7+1 | G_8 | G_9 | G_10 | G_11 |
|---|---|---|---|---|---|---|---|---|
| Lap 1 | male | 1.585 | 1.527 | 1.483 | 1.466 | 1.351 | 1.316 | 1.206 |
| | female | 1.067 | 1.034 | 1.015 | 0.986 | 0.963 | 0.929 | 0.904 |
| | total | 1.314 | 1.273 | 1.226 | 1.204 | 1.149 | 1.134 | 1.094 |
| Lap 2 | male | 1.446 | 1.459 | 1.407 | 1.393 | 1.326 | 1.286 | 1.286 |
| | female | 1.024 | 0.994 | 0.983 | 0.945 | 0.927 | 0.903 | 0.882 |
| | total | 1.265 | 1.224 | 1.185 | 1.146 | 1.128 | 1.104 | 1.051 |
| Lap 3 | male | 1.474 | 1.406 | 1.366 | 1.338 | 1.305 | 1.266 | 1.203 |
| | female | 1.006 | 0.943 | 0.962 | 0.917 | 0.886 | 0.863 | 0.816 |
| | total | 1.225 | 1.161 | 1.14 | 1.114 | 1.104 | 1.034 | 1.029 |
| Lap 4 | male | 1.406 | 1.378 | 1.353 | 1.306 | 1.272 | 1.227 | 1.157 |
| | female | 0.977 | 0.956 | 0.926 | 0.895 | 0.881 | 0.836 | 0.814 |
| | total | 1.186 | 1.119 | 1.107 | 1.074 | 1.082 | 1.046 | 0.868 |
| Lap 5 | male | 1.465 | 1.326 | 1.318 | 1.216 | 1.176 | 1.129 | 1.084 |
| | female | 0.914 | 0.895 | 0.866 | 0.869 | 0.853 | 0.826 | 0.763 |
| | total | 1.168 | 1.107 | 1.077 | 1.026 | 1.004 | 0.984 | 0.916 |
| Lap 1-Lap 5 | male | 1.477 | 1.414 | 1.389 | 1.345 | 1.283 | 1.243 | 1.187 |
| | female | 0.996 | 0.965 | 0.956 | 0.927 | 0.906 | 0.872 | 0.834 |
| | total | 1.204 | 1.174 | 1.147 | 1.106 | 1.082 | 1.055 | 1.016 |
| | rate of distance increase | 0% | 20% | 40% | 60% | 80% | 100% | 120% |
| | rate of male speed reduction | 0% | -4.27% | -6.22% | -9.50% | -14.42% | -18.24% | -23.33% |
| | Rate of female speed reduction | 0% | -3.11% | -4.15% | -7.22% | -9.71% | -13.69% | -18.58% |
| | Rate of overall speed reduction | 0% | -2.49% | -4.86% | -8.54% | -11.03% | -13.77% | -17.82% |

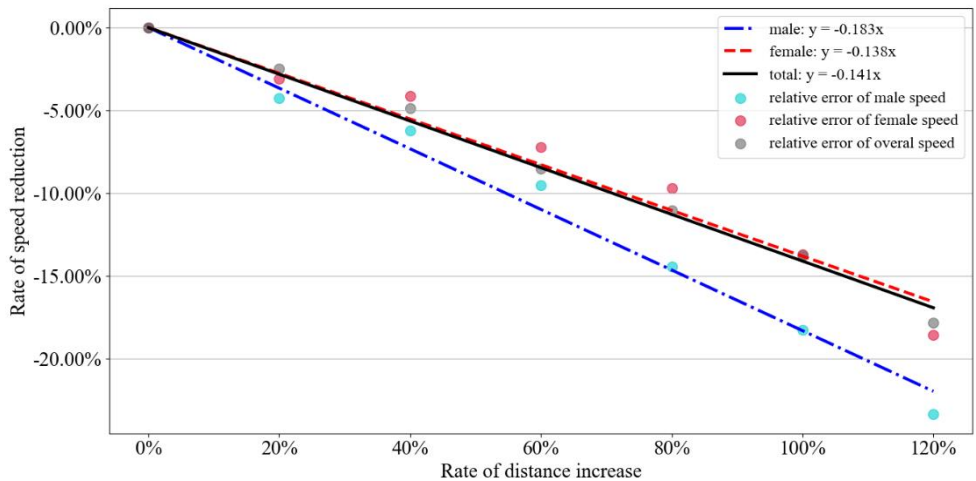

**Fig. 14.** The relation between distance increase and speed reduction.

## 4.4 BMI and speed

Prior to the experiment, the height and weight of the 42 subjects were recorded, which could be converted into the Body Mass Index (BMI). Based on BMI, there are three categories: underweight (less than 18.5), normal weight (between 18.5 and 23) and overweight (more than 23). As shown in Fig. 15(a), the speed of underweight participants is significantly faster than that of normal-weight and overweight pedestrians. As shown in Fig. 15(b), it is worth noting that there are no overweight girls in this experiment, underweight men move the fastest in the stairwells, followed by normal-weight men and overweight men, which is consistent with the trend shown in Fig.15 (a). In terms of gender, the male also moves faster than the female, and even the speed of overweight men is faster than that of underweight women.

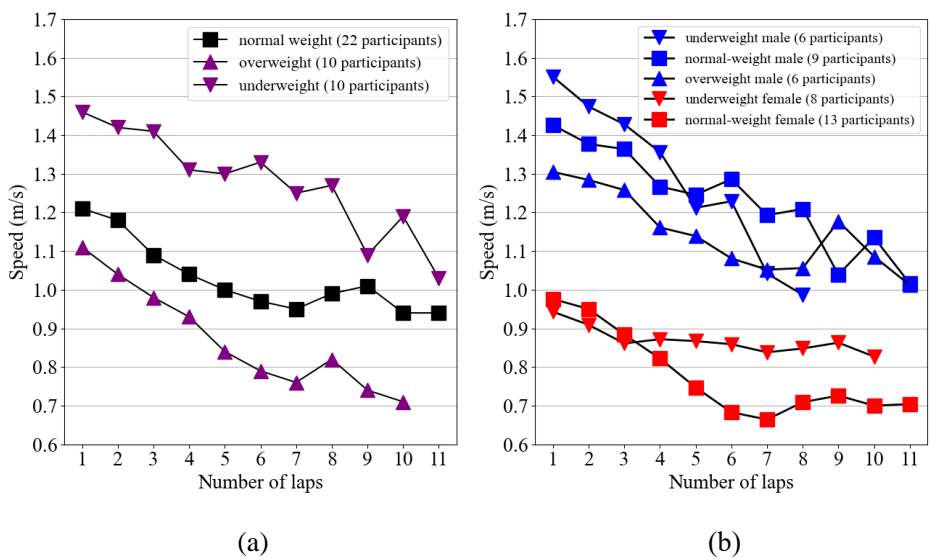

(a)          (b)

**Fig. 15.** The effect of BMI on evacuation speed on stairs. (a) Without considering gender. (b) Considering gender.

## 4.5 Heart rate

A turn in this experiment not only requires the participants to move downward from the 12$^{th}$ floor in the stairwells to the ground floor, but also includes taking the elevator back to the 12$^{th}$ floor. The bracelet worn by the pedestrian can record the change of heart rate with time during the whole evacuation. The heart rate of pedestrian increases when they descend in the stairwells. However, their heart rate decreases when they take the elevator to rise, this period of time can be considered as a rest period. In the whole process of evacuation, the change of heart rate was a process of wavy rise (Fig. 16).

In order to investigate the correlation between the speed and heart rate of pedestrians moving down on the stairwells, the speed and heart rate of male and female in each lap are shown in Fig. 17. With the increase of laps, the heart rate shows an increasing trend, and the heart rate of male is significantly higher than that of females. However, the speed of male and female decreases with the increase of laps, the speed and heart rate show a negative correlation with the increase of laps.

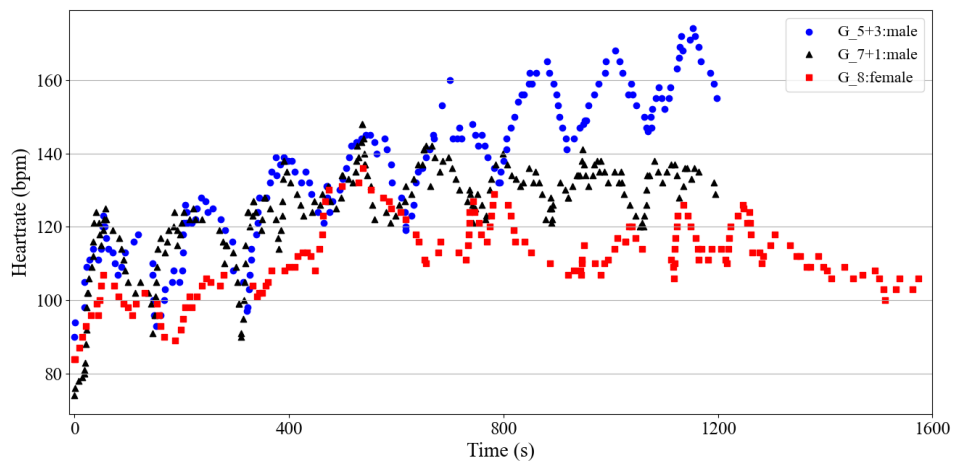

**Fig. 16.** Typical heart rate trends: taking the heart rate data of three randomly chosen pedestrians as an example.

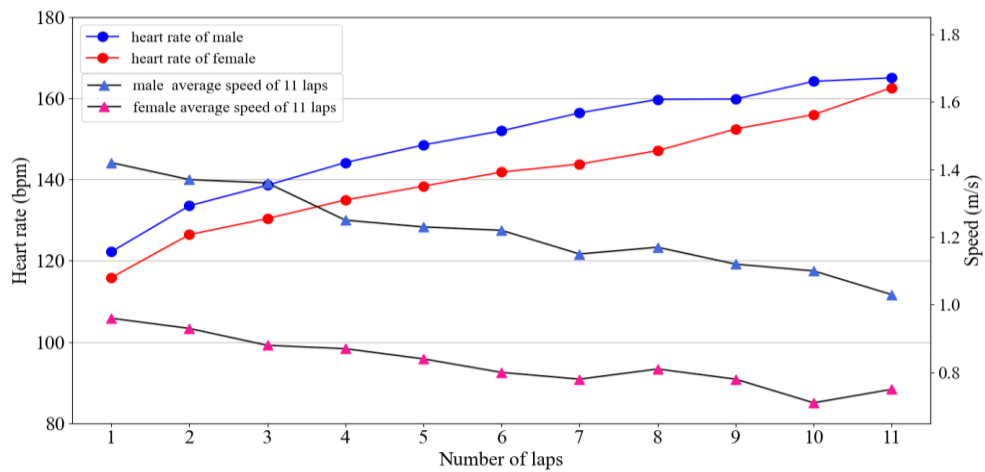

**Fig. 17.** The relations between heart rate and speed in each lap.

# 5. Discussion

Here we discuss the effects of influential factors on speed based on the experiment results in our study.

(1) Actual evacuation distance: As shown in Fig. 5 and Fig. 8, there is a great correlation between evacuation speed and evacuation distance. The greater the evacuation distance, the lower the pedestrian movement speed on the stairs.

(2) Expected evacuation distance: Pedestrians will set "appropriate speed" for themselves according to the evacuation distance. On one hand, since the expected evacuation distance is the shortest for G_5+3, we observe the maximal evacuation speed, and the expected evacuation distance for G_11 is the longest, the minimum evacuation speed can be found (Fig. 11- Fig. 13). As expected evacuation distance increases, the speed tends to decrease, as shown in Fig. 14. On the other hand, in fact, the psychological expectations of the G_5+3, G_6+2, and G_7+1 changed once during the evacuation. We find that the speed of the $6^{th}$ lap of G_5+3 has an upward trend relative to the $5^{th}$ lap; the speed of the $7^{th}$ lap of G_6+2 has an upward trend relative to the $6^{th}$ lap; the speed of the $8^{th}$ lap of G_7+1 has an upward trend in the $7^{th}$ lap, which is also reflected on males and females (Fig. 11-Fig. 13).

(3) Gender: On one hand, the evacuation speed of the female is obviously slower than that of the male in the evacuation. On the other hand, with the increase of evacuation distance, both male's and female's speed decreases, but the decline trend of male's speed is significantly sharper than that of female (Eq. (5) -Eq. (7)). This means that in the same scene, even if the speed of male is fast in the early period, as the evacuation distance increases, the fatigue will decrease the speed of men more quickly.

$$\text{first 8 laps}: \begin{cases} male: y = -0.039x + 1.447, x = 1,2,\ldots,8 \\ female: y = -0.024x + 0.969, x = 1,2,3,\ldots,8 \end{cases} \quad (5)$$

$$11 \text{ laps}: \begin{cases} male: y = -0.036x + 1.436, x = 1,2,\ldots,11 \\ female: y = -0.021x + 0.951, x = 1,2,3,\ldots,11 \end{cases} \quad (6)$$

$$\text{adjacent floors}: \begin{cases} male: y = -0.021x + 1.383, x = 1,2,\ldots,11 \\ female: y = -0.015x + 0.956, x = 1,2,3,\ldots,11 \end{cases} \quad (7)$$

(4) Having a rest: In this experiment, after a rest of $33.50 \pm 0.18$s (the time for pedestrians to rise from the ground floor to the $12^{th}$ floor by elevator), the local speed on the stairs from the $12^{th}$ floor to the $11^{th}$ floor will increase significantly (Table 4). In fact, the current theory of refuge story setting is not perfect, and the setting of refuge story in the evacuation strategy of super high-rise building is not clear. There are also no clear conclusions on how many floors should be separated by a refuge floor and how much rest time on the refugee floor is the best. There are even some super high-rise buildings that do not have refuge floors, which does not make sense from a security standpoint. The evacuation of pedestrians over a long distance will inevitably reduce the evacuation speed of pedestrians in the stairwell due to the physical loss. Therefore, in this paper, after pedestrians evacuate down 12 floors, which is like coming to the refuge floor. The time taken by elevator can be regarded as the time to rest on the refuge floor. From results in Table 4, we can draw the conclusion that the speed of pedestrians does increase after a rest on refuge floors.

(5) BMI: we study the evacuation speed of individuals with different BMI. For both male and female, underweight participants always have the fastest evacuation speed in this experiment, followed by normal-weight and overweight pedestrians (Fig. 15).

# 6. Conclusions

We conducted an evacuation experiment in Shanghai Tower in 2017, and the results showed that the evacuation speed of pedestrians moved from 117$^{th}$ floor is faster than that of pedestrians descended from 126$^{th}$ floor. Therefore, a hypothesis is proposed: pedestrians will set an "appropriate speed" according to the building height or evacuation distance of super-high buildings. In order to verify the validity of the hypothesis, an evacuation experiment was carried out in a 12-story building with a vertical distance of 40.920m. Participants of 7 groups (G_5+3, G_6+2, G_7+1, G_8, G_9, G_10 and G_11) were asked to descend in the stairwells from 12$^{th}$ floor to the ground floor and rise to the 12$^{th}$ floor by elevator. For certain group, such evacuation movement was repeated 8 to 11 times, so as to study the evacuation characteristics of a single pedestrian on the stairwell in a similar super high-rise building. The evacuation process was recorded by surveillance videos and 4092 local speed points between adjacent floors were extracted. The fitting formula of speed and the distance is proposed, as well as the method of quantifying the psychological expectation on the "appropriate speed".

The results show that the descending speed of pedestrians in the stairwells is related to evacuation distance, gender and psychological expectation. First of all, with the increase of evacuation floors, the speed decreases when running 8 to 11 laps in the experiment. Second, the evacuation speed of male is obviously faster than that of female, but the trend of decline of male speed is more obvious than that of female. Third, the evacuation speed will increase after pedestrians have a short rest in the elevator. Last but not least, according to the expected evacuation distance, pedestrians will set an "appropriate speed" in the whole evacuation process. If psychological expectations change during evacuation (G_5+3, G_6+2 and G_7+1), the evacuation speed will be improved temporarily.

Different from the previous studies, in this experiment, the maximum descending floor is 132 floors with a vertical distance of about 449.350m, and the maximum pedestrian travel distance on the stairwells is 1000.208m, which will refresh the record of pedestrian evacuation experiment of super high-rise buildings. In addition, heart rate of the pedestrian in the downward process is recorded through smart bracelet, and the correlation between speed and BMI and heart rate is analyzed.

The conclusion of this experiment is a supplement to the law of pedestrian evacuation in super high-rise buildings. The experimental results provide data and theoretical support for the setting of the refuge floor and the setting of the evacuation speed of the pedestrians in the stairwell of the super high-rise buildings in simulation models, and provide data support for the emergency evacuation of the super high-rise buildings.

# Acknowledgement

This work was supported by National Natural Science Foundation of China (72074149), and Program of Shanghai Science and Technology Committee (19QC1400900, 18DZ1201500).